# A Reanalysis of Eurasian Population History: Ancient DNA Evidence of Population Affinities

Casey C. Bennett (Department of Anthropology, Indiana University-Bloomington), Frederika A. Kaestle (Department of Anthropology and Institute of Molecular Biology, Indiana University-Bloomington)

Corresponding author:

Casey Bennett

Indiana University

Department of Anthropology

Current Mailing Address:

9806 Bonaventure Pl. #1

Louisville, KY 40219

Phone: (502) 384-5151

fax: (812) 855-4358

cabennet@indiana.edu

**Abstract**

Mitochondrial hypervariable region I genetic data from ancient populations at two sites from Asia, Linzi in Shandong (northern China) and Egyin Gol in Mongolia, were reanalyzed to detect population affinities. Data from a total of 51 modern populations were used to generate distance measures (Fst's) to the two ancient populations. The tests first analyzed relationships at the regional level, and then compiled the top regional matches for an overall comparison to the two probe populations. The reanalysis showed that the Egyin Gol and Linzi populations have clear distinctions in genetic affinity. The Egyin Gol population as a whole appears to bear close affinities with modern populations of northern East Asia. The Linzi population does seem to have some genetic affinities with the West as suggested by the original analysis, though the original attribution of "European-like" seems to be misleading. This study suggests that the Linzi individuals are potentially related to early Iranians, who are thought to have been widespread in parts of Central Eurasia and the steppe regions in the first millennium BC, though some significant admixture between a number of populations of varying origin cannot be ruled out. The study also examines the effect of sequence length on this type of genetic data analysis and provides analysis and explanation for the results of previous studies on the Linzi sample as compared to this one.

**Introduction**

Recent analyses of ancient DNA from sites in northern China and Mongolia have provided interesting results regarding the genetic history of the region and of eastern Central Eurasia in general (e.g. Wang et al. 2000, Keyser-Tracqui et al. 2003). The period of the sites in question stretches from around the middle first millennium BC to the first few centuries AD and represents an important time period in the area: the rise of the Han dynasty in China and the Hsiung-Nu on the Mongolic steppe, the possible earliest appearances of Turks and Mongols, and the earliest attested conflicts between ancient Chinese and steppe peoples of Inner Asia. Elsewhere in Central Eurasia, the Scythians and Sarmatians appeared in the farthest western portion of the steppe in south Russia and have been putatively connected to Indo-Iranians or Iranians (a branch of Indo-European). In the central portions of the steppe (roughly modern-day Kazakhstan) not much is known for certain, though there is evidence of a group(s) of people referred to as the Saka, who are commonly identified as Indo-Iranian (or Iranian) and were nomadic pastoralists like the Scythians and Hsiung-Nu. More highly attested are the Sogdians, sedentary Iranians of the Transoxus region. Further east, in present day Xinjiang, there were possibly Indo-European peoples such as the Tokharians (a group of Indo-European speakers attested with recorded documents) and the peoples represented by the various mummified remains from the region in the second millennium BC through the first few centuries AD. Along with these peoples there are of course many others of whom we know very little or nothing in this period (e.g. Ob-Ugrians) (for a general discussion of the above, see Mallory 1989, Sinor 1991, Mair 1998). The two ancient sites in this study, Egyin Gol (Keyser-Tracqui et al. 2003) and Linzi (Wang et al. 2000), thus reflect a key period in the region (as well as Central Eurasia in general). It is clear that changes in the social, political, economic, and cultural realms occurred.

However, the exact degree to which these various cultural and linguistic groups represented biological populations is debatable, and it is unclear whether the aforementioned changes were accompanied by the movements of such biological populations.

The question of the ancient history of northern China and Mongolia is a difficult issue. Traditionally, many have taken the approach that ‘China is an island.’ However, this invariably is false (as with the ‘Europe is an island’ model). Connections existed across Eurasia back to at least the first millennium, if not earlier (Bentley 2000). Moreover the connection of the biological past to the cultural past has not been clearly detailed, although various arguments have been made. Lattimore (1951) suggests that the difference between the peoples of Central Asia (which he defined as Manchuria, Mongolia, Tibet, and Chinese Turkestan) and those of sedentary China (which he defined as the primarily agricultural areas of China proper) was the difference between an extensive pastoral economy in Central Asia (although there are some places with agriculture or a mixture of economies including Manchuria and the oases of Sinkiang) and an intensive agricultural economy in China. He also points to the inability of states with a mixed economy of both pastoral nomadism and intensive agriculture to succeed (though this is not entirely true, case in point Manchuria or historical “Central Asia”). Lattimore further suggests that the “Northern Barbarians” were originally of the same ethnic stock as Northern Chinese but were split through economic differentiation. This led to differentiation in the rates of change (of culture, technology, etc.) that split these early peoples into two “orbits.” Lattimore argues that it was the expansion of the early Chinese that pushed out the peripheral groups who would become the early “barbarians” by the 5th century BC. However, the question of connections between the peoples of the steppe has continued to generate a large amount of work, some of which conflicts with Lattimore. A.P. Okladnikov (1990) argued that early on there were Europoid peoples in Inner Asia who would later move down off the steppe into India and Iran (and Europe as well). According to this view, Mongoloids, who are traditionally thought of as inhabiting the Inner Asian areas, did not appear until around 1200 BC. Further complicating the question of the human biological history of eastern Central Eurasia are the so-called "mummies of Urumchi" or "mummies of the Tarim Basin," who have often been associated more with "Europoids" or "Caucasoids" rather than "Mongoloids". These remains have not only been potentially related to Indo-Europeans in a biological sense, but also culturally (e.g. "Tartan" clothing). They have also been putatively connected to various Indo-European groups of multiple time periods from around the region, including the Tokharians, the Saka, the Andronovo of the Central Asian steppe, as well as the Afanasievo of the Altai and western Sayan ranges in southern Siberia (Mair 1995, 1998).

Adding to this debate are the various theories of Indo-European origins and expansions, as argued by both Mallory (1989) and Renfrew (1987) as well as numerous others. The main component of this theory is that the Indo-Europeans originally represented a centralized cultural group, though there is great debate over the location of their origins and the time of dispersal and expansion as well as possible routes. Though Renfrew (1987) has argued for an Anatolian origin connected to the spread of Neolithic farming, there is an alternative argument detailed by Mallory, Gimbutas, and others (see Mallory 1989), who connect the Indo-Europeans to the south Russian steppe, possibly around the Black and/or Caspian sea as well as the southern Urals or northern Caucasus (and/or possibly Eastern Europe). The exact ranges of this "homeland" are debatable, though estimates have been given. Also, the exact correlation and geographical positioning between the different groups of Indo-European speakers before dispersal is not clear. Indo-European languages are generally divided up into centum (European, western) and satem (Indo-Iranian and Indo-Aryan, eastern) languages, though some discrepancies such as Tokharian (a centum language in the east) do exist (Mallory 1989).

The earliest expansions of Indo-Europeans are generally dated to sometime between the 5$^{th}$ millennium BC and the 3$^{rd}$ or 2$^{nd}$ millennium BC. These expansions have been connected by Anthony (1995) to the domestication of the horse on the steppe and to the later development of the war chariot (as well as wheeled vehicles in general, metallurgical developments, and herding). Some of the earliest evidence of horse domestication is at the site of Dereivka in the south Russian steppes dated to around 4000 BC, which is connected to the Stredni Stog culture (Anthony 1995). This evidence is related to possible bit wear, though see Levine (1999) for dispute. However, there is evidence that men may have hunted horses (as well as other animals) on the southern portion of the steppe as early as the late Paleolithic (Praslov 1989), suggesting that man may have had long contact in the region with horses. Also, some of the earliest evidence of chariots found to date comes from the Sintashta-Petrovka culture (possibly related to the Andronovo) on the steppe near the Volga-Caspian region and the Urals, dated to around 2000 BC (Anthony 1995). Early possible expansions of Indo-Europeans include the Germanic peoples, Celts, Greeks, Latins, and others into Europe as well as expansions east such as the Andronovo culture (Mallory 1989). The earliest eastern expansion may have been the aforementioned Afanasievo culture in the mid 4$^{th}$ millennium BC (Anthony 1998). Other movements include the possible migrations of Indo-Europeans into Xinjiang at least as early as the early second millennium BC (Kuzmina 1998) and the historically attested movements of Indo-Iranians and Indo-Aryans south into the Iranian Plateau region and India, though the exact nature or sequence of these is not certain (Mallory 1989, Parpola 1998). The connections of these eastern peoples of the putative Indo-European family farther east, such as into China, is subject to much scholarly debate, though there is some evidence of Indo-European loan words in Old Chinese as well as cultural and technological changes in northern China in the 3rd and 2nd millennium BC (Pulleyblank 1996, Kuzmina 1998, Beckwith 2002, Di Cosmo 2002). Certainly, sites such as Zhukaigou (roughly 2000 BC, Linduff 1995) and Linzi (Liangchun site, roughly 500 BC, Wang et al. 2000) in northern China, as well as mummies of the eastern Tarim Basin suggest that the history of the region, both culturally and biologically, may be very different from what it is today or even in known historical times. These may indicate alterations in the biological and cultural makeup of the region occurring as early as the Bronze Age or late Neolithic and even possibly earlier; though obviously because of the often poor connection between culture and biology, the relationships between the two must be examined with a fair bit of caution. Other sites, such as Egyin Gol in Mongolia (Keyser-Tracqui et al. 2003), shed light on the later shifts in the region, as well as explore the possible connections between the differentiation of the steppe peoples (or lack thereof), in their early stages of development, and China.

The purpose of this particular study is a reexamination of two sites from eastern Central Eurasia, Linzi in China (Liangchun site) dated around 2500 years before present (Wang et al. 2000), and Egyin Gol in Mongolia dated to around the last few centuries BC to the first few centuries AD (Keyser-Tracqui et al. 2003). Linzi is in Shandong province in northern China, near the Yellow river and the Ordos region. It is presently part of the city of Zibo. Sixty-three individuals were examined in the original study from the Liangchun site in Linzi dating to around 500 BC. However, only 34 gave good results for mitochondrial DNA. Though the original study extracted longer sequences (287 bp), only the 185 bp segments (nt 16194-16378) actually used in their analysis were available in GenBank. The date places the material during the Spring-and-Autumn period in between the fall of the Eastern Chou dynasty and the rise of the Han dynasty. The original study also included 50 modern Han Chinese individuals from Linzi (labeled Qidu here). Samples also have been examined from the more recent Yixi site at Linzi (2000 before present, Oota et al. 1999), though these were not included here due to inconsistency in the Yixi data in GenBank (nearly half lacked sufficient data for inclusion). Because the length of the sequences available is only 185 base pairs, accurate comparative genetic analysis with other populations is difficult. The authors report that they found the Linzi material clustered closely with modern Europeans, particularly Finnish, Turkish, and Icelanders (Wang et al. 2000). However, as we will show, this analysis may be imprecise (as suggested by Yao et al. 2003). Additionally, we will examine the fact that the putative “cousins” of the Europeans, the so-called Indo-Iranians, are known to have been widespread in Central Eurasia at that time.

Egyin Gol is a necropolis in northern Mongolia (labeled simply Egyin below). The original study successfully extracted DNA successfully from 62 specimens ranging from around the 3rd century BC to the 2nd century AD, including mitochondrial and nuclear DNA. In this study, we will examine the mitochondrial DNA (nt 16009-16390). The site sits along the Egyin Gol River, a tributary of the Selenge River, which flows into Lake Baikal. The site has been attributed possibly to the Hsiung-Nu, who the authors describe as an ancient “Turkomongolian” tribe (Keyser-Tracqi et al. 2003). However, the exact relationship of the Hsiung-Nu to either Mongols or Turks (who do not definitively appear until the first few centuries AD with the Jou-Jan and their “blacksmith slaves” the Turks, (Sinor 1990)) is not clear. Moreover, the Chinese records of the Hsiung-Nu have proven them difficult to classify culturally and linguistically, and the origins of the Hsiung-Nu are not clear (Di Cosmo 2002). The necropolis was divided into three sectors (there was also a fourth zone, D, but no DNA samples originate there), with A (the oldest) and B representing older sections, after whose fusion sometime in the early centuries AD new graves were dug in what is called sector C. The authors also point out that both the paternal lineage and the mtDNA sequences shared by four of the paternal relatives in sector C have been found in modern day Turkish individuals, as well as in two of the graves from the older A and B sectors. The authors suggest that this evidence may point to a “Turkish component” to the Hsiung-Nu tribe in later periods (Keyser-Tracqui et al. 2003). In this study, we will attempt to examine the affiliation of these individuals to populations around Eurasia, as well as to look at any possible differences in the genetic relationships of the older A and B sectors to the newer possibly “Turkic” sector C.

**Materials and Methods**

For the purposes of this study, we examined as wide a range of variation as possible at the population level. For this, we compiled 3,703 mtDNA HVS I sequences from 51 modern populations (including the Qidu samples) from across Eurasia in addition to the two aforementioned ancient populations (total populations = 53). Most of the populations were included regardless of whether they were thought to be related to the two ancient populations, though a few populations were added because of hypothetical relationships, and the African Biaka were included as an outgroup. Limitations in computer power and software design placed restrictions on the total number of populations employed in the analysis, but populations were included up to reasonable limits of this constraint. For example, DNAsp, a program used in the analysis, consistently failed at around 2500 sequences or over (other programs, like Arlequin, are written specifically to handle a maximum amount, in Arlequin's case that amount is 1000 sequences). Further, any increase in populations or sequences increased the number of calculations needed at an exponential rate. Even the use of the local supercomputing network could only ameliorate these issues, not eliminate them. Thus, the idea here was to minimize bias within the constructed sample. Although additional populations, such as Tibetans, Russians, and some Siberian groups, are available, the analysis of such a large dataset proved prohibitive.

The populations included are listed in Table 1. The table includes individual population data for several values, as well as the values for the total sample with and without Linzi and Qidu (in order to account for the short sequence length of Linzi and Qidu, which may result in inaccurate values when included). Note that the Basque data is not included in Table 1 for reasons specified below. The averages for the values within populations are also included. The average number of individuals sequenced was 71, though this varied widely, along with an average of 47 haplotypes per population, excluding all gaps or missing data. The average number of differences between sequences within populations was 5.73, with an average nucleotide diversity (the average number of differences per site between any two sequences) of roughly 0.015 and an average haplotype diversity (a measure of genic variation, which equals $1 - \Sigma x^2$, where x equals the haplotype frequencies) of .96. However, individual population averages appear to be dependent to some degree on sample size and sequence length, as would be expected. For average number of differences, there was a significant correlation to sequence length (.571, $\rho = .000$) but not sample size (-.263, $\rho = .058$). For nucleotide diversity, there was a significant correlation to both sequence length (-.438, $\rho = .001$) and sample size (.290, $\rho = .035$). Note that the diversity indices for the ancient populations fall within the range of those of modern populations, which indicates that they are comparable to modern populations in terms of variation.

These populations were divided into three loosely defined regions at the discretion of the authors. These groups were: Europe (Armenians, Georgians, Mari, Moksha, Saami, Slovakians, RomB, RomS, Rom2S, Germans, Hungarians, Cumans, Basques, Catalans, Icelanders, and the Moroccans), South and Southwest Asia (Lambadi, Lobana, Uttar Pradesh, Boqsa, Pushtoon, Pakistan, Parsi, Iranians, Iraqi, Kurds, Turks, Kashmir, and Tunisians), and East and Central Asians (Kazakh, KazakhXJ, Uighur, UighurXJ, KirghizHL, KirghizLL, Guangdong, Guangdong2, Yunnan, Vietnamese, Indonesians, Akha, Koreans, Japanese, Mongolians, Ewenki, Wuhan, Shandong, Liaoning, and Qidu). Though there could be some debate over the division into these regions, it is not extremely important since, as we will see below, only the top 9 or 10 populations from each region were used for the final comparison. These initial regional divisions were necessary in order to break up the data into manageable datasets. Moreover, the purpose of the study was not to determine the vast connections between populations in Eurasia, nor do we claim that the results can be used in this way. Rather, the purpose is a very narrow focus, that being how these modern populations from various regions relate to the two ancient populations of Linzi and Egyin Gol. To this purpose, we calculated regional distance measures (principally Fst's, a measure rooted in heterozygosity values within and among populations) for each region and then included the top nine or ten "matches" (the lowest Fst's) from the region for a "total" comparison. The central assumption here is that the populations from each region best represent that region as far as biological relation to the two ancient sites. We may lose in this method of comparison some of the minutiae of the more distant relations, but the closer relations (such as the top ten) should be accurate. Obviously, a particular Central Asian population might have seemed relatively closer to the Egyin material if placed in the South and Southwest Asian region, for instance. However, every population was given a fair chance to compete ("free competition") instead of arbitrarily being included or excluded in the analysis, which we believe results in a more accurate estimate of relationships.

A quick glance over the regional populations is probably in order (see the original sources for more detailed information). The Europeans included Germans and Icelanders representing the Germanic and Scandinavian peoples. The Catalans represented a Western European population, while the Basque represented a supposed isolate in Western Europe. The Slovakians represented East Europeans and Slavs. The Hungarians are Ob-Ugrian speakers, though they probably have at least some historical connection to Turks (the name Hungarian probably derives from Onoghur, a Turkic people of Central Eurasia in the early middle ages, Golden [1991]). The Cumans were originally Turks known by various names in different sources. The Rom populations are gypsies of Eastern Europe, whom many believe to be ultimately descended from northern Indian ancestors (Gresham et al. 2001). The Saami are Finnic (Ob-Ugrian) speakers from northern Scandinavia, while the Mari and Moksha are Ob-Ugrian speakers from Russia closer to the Urals. The Armenians and Georgians represent Caucasus populations. The Moroccans were included here as a check against suggestions of significant gene flow across the Mediterranean between North Africa and Southern Europe, at least in the West (Plaza et al. 2003).

The South and Southwest Asians included the Uttar Pradesh and Boqsa samples from Uttar Pradesh. The Lobana are also northern Indians from Punjab. The Parsi are from northwest India (mostly Gujarat) but supposedly descend from Iranian migrants (hence the name, Pars, Fars, Persians). The Lambadi are the "gypsies" of India, from whom the gypsies of the world are hypothetically descended (see above). However, most Lambadi live in the North and Northwest of the Indian subcontinent, while these samples come from Andhra Pradesh in the East Central region. The Pakistan and Pushtoon (original source name preserved, likely Pashtun) are from Pakistan, while the Kashmiri are from Kashmir. Further east, we have the populations of Iranians, Iraqi, and Turks representing the Near East, as well as the Kurds from Iraq. The Tunisians were also included as a related North African population that may have absorbed similar Arab or Near Eastern gene flow since the genesis of Islam.

In East and Central Asia, the Kazakhs, KirghizLL (lowland), KirghizHL (highland), Uighurs, UighursXJ (Xinjiang), and KazakhXJ (Xinjiang) represent some of the diversity seen in Central Asia and Xinjiang today. The Mongolians, Ewenki, Koreans, and Japanese along with the northern Chinese populations of Liaoning, Qidu, and Shandong represent the northern part of East Asia save Siberia. The Yunnan, Guangdong, and Guangdong2 populations represent the southern part of China. The Vietnamese are included to have some comparison for Southeast

Asia, though according to the original reference they are actually “first generation immigrants” to California. The Akha are tribal people from Thailand, who along with the Indonesians and Vietnamese should represent Southeast Asia. The Wuhan province is fairly centrally located in China, while the Xinjiang Han should represent ethnic Han Chinese in the far western reaches of China.

Some population sequence data was collected from GenBank (Benson et al. 2000) and HVRBase (Handt et al. 1998). Other population sequence data was generated from the literature or from a data table provided by Toomas Kivisild and Mait Metspalu (personal communication 2003), using a program specifically written for large number sequence creation (from lists of nucleotide differences) by one of the authors (CB). Sequences were either initially aligned using the Sequencher program (Gene Codes Corp.), or were automatically set up to be aligned if created using the aforementioned program. Any sequence format conversion was handled by a program written for large number sequence conversion by one of the authors (CB). After initial alignment and creation, sequences were imported into MacClade (Maddison and Maddison 1989) for final alignment and editing purposes. All of the sequences were edited to extend from nt 16001-16497, either by cutting longer sequences or inserting “n’s” in shorter sequences. There are a few quick notes that should be made. The Basque population was downloaded and discovered to contain the same sample label for several samples. Because it was unclear whether these represented the original 45 individuals from the study with some identical sequences or sequences from 27 individuals with duplicates, duplicates were edited out. Thus the Basque sample may represent all of the variability taken from the original study, but not the proper frequencies. As mentioned above, they are included on the data table but their population values are not, and they were not included in total calculations or averages. Also, several sequences had to be removed from both the Slovakian and Rom populations taken from GenBank, because it was not clear what they represented (certainly not the HVS I, perhaps the HVS II?). Also, the Pushtoons were reduced to 360 base pairs long (nt 16024-16383) because of some confusion over primer lengths. The resulting sequence set went through a final round of alignment by hand editing. Note that any inserted deletions in the cytosine tract were removed by consensus (since this was the standard in 51 out of 53 of the populations originally). It was assumed that it is not clear whether the alterations in the cytosine tract of some sequences are deletions and extensions of the cytosines, or transversions of bases in the preceding poly-A segment to cytosines. This was only done in the Korean sample, since all other sample sets were apparently aligned this way originally. Also note that there was an insertion of a deletion in most sequences at the end of the cytosine tract to account for an extension of the cytosine tract in some individuals by one (making a total of 15 bases in the cytosine tract and preceding poly-A segment rather than 14), which was found in one Lobana, one Vietnamese, one Egyin, and several Icelanders. Along with another insertion at 16104 (16104a), one in a later polycytosine segment (16262a) and the aforementioned insertion in the cytosine tracts (16194a?), the final sequences including all unknowns and gaps were 500 nucleotides long.

After the final round of alignment, the sequences were analyzed using DNAsp 3.99 (www.ub.es/dnasp; Rozas and Rozas 1999). This program was used to generate distance data for the various regional group tests and for the total tests. Separate tests were run with Linzi (and Qidu in the case of East Asia), since these sequences are only 185 bp, and the Egyin material (excluding Qidu in the case of East Asia), with sequences of 382 bp long. The Egyin material was included in the Linzi runs for comparison purposes. The analysis below on the effect of sequence length on this particular method, though, does suggest that it should be minimal. Each run included only the sites for which we had data across the entire dataset of that run. Due to the variation in the populations included, each run utilized slightly different sequence lengths, since various sequences and populations had differing amounts and locations of missing data. There was also a final third run with the Egyin population split to test the possibility suggested by Keyser-Tracqi et al. (2003) of a “Turkic” component in the Egyin sector C material, in which the Egyin material was divided up into EgyinAB (from sectors A and B) and EgyinC (from sector C) as described by Keyser-Tracqi et al. (2003) (results not shown). As previously mentioned, the Biaka population was included in every run as an outgroup.

Three distances matrices were generated for each run: Fst’s (Hudson et al. 1992), Nst’s (Lynch and Crease 1990), and Da’s (Nei 1987). The main analysis centered around the use of Fst’s to estimate distances (which generally reflected all of the distance measures generated, see below). Each run (Linzi, Egyin, and Egyin Split) was done for each region. After the regional analysis for each of the three runs, the top matches (those populations with the lowest Fst’s relative to the probe population of that run, Linzi or Egyin) were selected out of each region for a composite analysis. In the case of the first two runs, the top ten from East and Central Asia, the top nine from Europe, and the top eight from South and Southwest Asia were chosen. This was because of differences in the number of populations for each region (East and Central Asia with nineteen populations, Europe with sixteen, and South and Southwest Asia with thirteen). A different approach was used for the Egyin split runs, since to have the exact same set of populations for a single total run for comparative purposes, compromise sets of the top matches were done.

The total runs were completely recalculated, starting over by recalculating new distances for the new “global” (or more properly Eurasian) population sets using DNAsp. The same procedure was followed as above using multiple distance measures. All of the distance data tables below are generated from Fst's. However, several tests were done with the other two measurements, and they were found to follow the same general relative order of population distances (data not shown). Also, tests were done excluding the Basques from the European sets (since the Basque have been noted to have discrepancies) and no change was found in the relative order of the results. Note however that this does not mean that the Basque distance measurements are accurate, merely that tests were done to see if their inclusion or exclusion caused error in the rest of the results. It should also be noted that the applicability of Fst measurements to population comparisons is a highly debated issue, as are the problems associated with appropriate Fst calculation (Nei 1977 and 1986, Weir and Cockerham 1984, Long and Kittles 2003).

Haplogrouping was not done in this study. Whereas there is no estimate for the number of haplotypes in the total study (due to analytical problems associated with the large dataset), an estimate gathered from sequences just 217 bp in length and excluding all missing sites as well as Linzi and Qidu found 1084 haplotypes. If all the data were included, this number would likely increase. Although many of these haplotypes might cluster into haplogroups, analysis at this level would result in a loss of information. In addition, we lack sufficient sequence data for accurate haplogrouping in many cases (e.g. haplogrouping from 185 bp sequences with no restriction site data is difficult) and variability in sequence length would affect our results. Some information on haplogroups in the ancient populations is available in the original articles (Wang et al. 2000, Keyser-Tracqui et al. 2003), as well as Yao et al. 2003 (see discussion section below).

**Results**

This section will be broken up into several components, looking at the two initial runs by region and total followed by the Egyin split run. Note however that since the top matches for Egyin and Linzi were generated from their respective runs, the populations in their total population comparisons vary to some degree. Also, these data represent only part of the full matrix, and thus relationships between modern populations should not be inferred. The Egyin test is also one population shorter at the regional level since the Linzi data was not included in its run for reasons explained above. All distance data tables are sorted in order of lowest (closest) to highest (farthest) Fst values.

The Linzi and Egyin data were compared to European populations. The results from these runs are shown in Table 2. As can be seen in the data tables, consistent estimates across different runs and population sets are noted. However, the order of the relationships should be taken generally. For instance, the details of whether the Armenians are really .0048 closer to the Linzi samples than the Catalans are not really necessary for this study. What is important is to see the general relative order; that certain populations are near the top, others are in the middle, and still others are at the bottom.

We can see several interesting things in Table 2. First, the Linzi material seems to be closer to the European populations than to the Egyin individuals, except for the RomS sample. However, the calculated distances between ancient populations seem to be consistently greater than those between modern populations or between modern and ancient populations, perhaps reflecting increases in modern population sizes and/or gene flow (as to the effects of effective population size on Fst's, see Cavalli-Sforza et al. 1994), though this would not necessarily affect the accuracy of the calculations themselves (Holsinger and Mason-Gamer 1996). Another possible explanation is that there was nonrandom fission (such as along familial or clan lines) in ancient populations in contrast to the larger social units of modern populations which are less dependent on familial relationships (Smouse et al. 1981, Whitlock 1994). As to the RomS population, it includes only 16 individuals who cluster into only 2 haplotypes, which might have resulted in inaccurate associations. Either way, the other Slovakian Rom population (Rom2S) is not in the top ten for either Linzi or Egyin. We can also see in this data table that the Egyin do have some affinity to the Bulgarian Rom and the Moroccans, however this is only relative within the European populations, as we will see below. We note that the Icelanders are near the top of the Linzi list (as Wang et al. 2000 suggested), but several populations are closer, including the Hungarians at the top.

Table 3 contains the comparisons of the Linzi and Egyin samples to the South and Southwest Asians. There are several interesting results. First, we once again see that the Linzi individuals bear a closer affinity with most of the modern populations than with Egyin individuals except of course for the sub-Saharan African Biaka (though see above). Secondly, there is a definite difference between the two ancient populations in the ordering of this table. The Egyin list has the populations of India mostly at the top (save maybe for the Uttar Pradesh, which is still in the top half) with the Pakistani populations and the Tunisians (which upon further review were found to bear relatively closer Fst's with the Pakistani populations than with the populations of the Near East, possibly reflecting the Arab expansions). The populations of the Near East (Iranians, Iraqis, Kurds, and Turks) are all at the bottom. In the Linzi list, we see the opposite trend, with the populations of the Near East mainly at the top. The populations of Pakistan and Tunisia are mixed in the middle, while the Indian populations are all at the bottom. The top matches in this list seem to be the Iranians and Turks of Turkey. As to why the Egyin material bears a close affinity with the Indian populations, relatively speaking, in this list, this may have something to do with some degree of shared maternal heritage in South and East Asians dating back to the earliest settlements of South and East Asia, though this shared heritage is probably limited due to subsequent divergence in modern day populations (Kivisild et al. 2003). As we will see below, the Indians are probably just the closest matches from this region, not necessarily close overall. Further, this may relate to the affinity of Egyin and Linzi to East Asians more than to their direct affinity to Indians.

Table 4 contains the comparisons of Linzi and Egyin to East and Central Asia. Note here that Qidu was also removed from the Egyin run because it is only 185 bp long. The data table shows a differential clustering of Linzi and Egyin with these populations. Once again, we see that the Linzi material is actually closer to the modern Asian groups in this study than to Egyin (though see above). The top half of the Linzi list is dominated by Southeast Asians, southern Chinese, and Central Asians. The lower half is dominated by northern Asians (save for the lowland Kirghiz and Akha). As to how and why both Southeast Asians (and southern Chinese) and Central Asians are similar to the Linzi population, it is not clear, though this is only a relative comparison within this region. However, there is some debate over the nature of ancient East Asian genetic history, so possibly there are issues here that have yet to be illuminated (Yao et al. 2002b, 2002c, 2003, Oota et al. 2002). Further, there are some issues with the Vietnamese sequences (see below). Also, it should be noted that the modern Qidu samples, from the same general locale as ancient Linzi, were in the lower half of the table. The Egyin list shows the opposite trend. The top half of the Egyin list is dominated by northern East Asians (including northern Chinese) except for the Xinjiang Han (however, a closer analysis of the Xinjiang Han shows them to have a genetic affinity to both Central Asians and northern Chinese and Mongolians). The lower half of the list is dominated by Central Asians and Southeast Asians (save for the lowland Kirghiz, which appropriately do an exact reversal from the Linzi list by showing up as the closest Central Asian population), with the Southeast Asians mainly at the very bottom except for Guangdong2. The Wuhan sample from central China seems to float about in the middle of both lists.

Table 5 is the list for the total comparison of modern populations to Linzi and Egyin from a composite of the top matches from each region as explained in the methods section. Once again, it should be noted that the group of populations for each list is somewhat different and was generated independently from separate runs of Linzi and Egyin data in regional models.

The table clearly shows a differential pattern in the genetic relationships of Linzi and Egyin to other populations. First, we see that the Egyin and Linzi populations did not share a close affinity with each other, or at least not more so than they do with modern populations (though see above). As for the Linzi individuals, they seem to be most highly related to Near Easterners (Turks, Iranians, and Iraqis), Armenians, and eastern Europeans (Slavs, Hungarians), though others such as Catalans and Iraqis are mixed in. The Icelanders are twelfth on this list. The high placement of the Vietnamese may be an anomaly, error, or some element of ancient genetic history that is not clear (though see Yao et al. 2003). However, it should be noted that the Vietnamese sequences lack a section of bases near the cytosine tract, compounded by the large number of sequences compared in this population set, which could provide for some anomalous results. Furthermore, this approach cannot accurately account for significant admixture (a distinct possibility given the proposed haplotypes of some of the individuals at Linzi, see Yao et al. 2003 and the discussion section of this paper), though neither of the other Southeast Asian populations (the Akha from Thailand or the Indonesians) even made it into the composite run. Thus, given these issues, it should be reiterated that only general trends should be drawn from this study.

What is clear is that the Linzi material does have an affinity to the west, most highly to the groups mentioned above. The East Asians that made the list are generally toward the bottom, save for the Vietnamese. The other interesting thing is that the few Central Asian Turkic peoples are generally toward the bottom, with only the Uighur appearing in the middle of the top half (but still outside the top ten). It has been noted that Near Eastern Turks actually bear more affinity with Europeans and Near Easterners than with their linguistic cousins in Central Asia, and that the Turks came to dominate Turkey through an *elite dominance* process, meaning that the effect on the maternal heritage should be minimal (Comas et al. 1996, 1998). Thus we may be able to include them together with the Iranians and other Near Easterners, who bear a close affinity with Linzi, though the relatively high distance between the ancient Linzi sample and Central Asian Turks may actually be from more recent East Asian admixture. The other high affinity groups, mostly from Eastern Europe in the Slovakians and Hungarians, may be related either directly or through the indirect process of East-West settlement in Central Eurasia that has been occurring in Eastern Europe for at least the past several thousand years, beginning possibly with the Indo-Europeans and definitely by the time of the Iranian Scythians and Sarmatians, as well as with later Turkic groups (though we have noted the distance between modern Central Asian Turkic peoples and Linzi).

The Egyin list is, of course, from a much longer sequence comparison, thus increasing the probability of valid connections. Other than some reordering, the top matches are all the same ones from the East Asia regional table (Table 4). The middle of the list includes all of the South and Southwest Asians in the same general order as found in the regional comparison, while the bottom of the list includes all of the European populations, in a similar order. Note that there are significant gaps in the fall of the Fst's between the Europeans and the South and Southwest Asians and between the South and Southwest Asians and East Asians (except for the Lambadi and the RomB, who seem to float in between) of about .02 to .03, whereas no other populations *within* regions (except for the Lambadi and the RomB of course) exhibit gaps of even .01. However, note that this is not indicative of the difference between these groups of populations, but between these populations and Egyin. Further, it should be noted that this list does not specify exactly where non-included populations from the regional comparisons would fall. What is clear here is that the Egyin population seems to firmly relate to East Asian populations, particularly the northern East Asian populations (northern Chinese, Inner Mongolian Ewenki, Mongolians, Koreans, Japanese, and the Xinjiang Han whose partial northern East affinities were explained above). Whatever the exact interpretation of the genetic affinities of the Egyin and Linzi populations may be, it is clear that they differ significantly.

A further test was undertaken in this study to examine the suggestion of Keyser-Tracqui et al. (2003) about the possible differences of sectors A and B compared to sector C at the Egyin Gol site (with sector C showing "Turkic" affinities). It is questionable whether it is reasonable to pursue such an examination from a methodological standpoint. The division of the population here into subpopulations is based on observed variation in spatial and temporal factors at the necropolis, as well as some putative differences in genetics. However, the differences in these subpopulations based on these factors may be only superficial differences, and the division thus arbitrary in nature. Further, the differences in genetics may not be reliable with two subpopulations of size 38 (EgyinAB) and 8 (EgyinC). Despite this, the test was done tentatively to see if there were any significant differences in affinity to modern populations or to each other.

The final results for both supposed subsamples generally followed along the lines of the total Egyin population analysis, and there were no clear distinctions between the two groups (though there were a few slight differences). However, given the methodological issues, this part of the study is not included in this paper, and the results are not shown.

We also examined the suggestion by Yao et al. 2003 that the results from Linzi were due purely to the short sequence length. For this, we took the Fst distance data from the Egyin total run for Egyin (318 bp) and compared it to the calculate Fst values when Linzi and Qidu were added (166 bp, referred to as Egyin Limited or EgLim, see table 6). We also took the Japanese Fst data from the Linzi total run (with the Japanese and Liaoning added, 149 bp, referred to as Japanese Limited, or JpLim) and compared it the calculated Fst values when Linzi and the Vietnamese were removed (319 bp, table 7). Though there is some slight movement of populations in both cases, there is no major discrepancy in the relative ordering of the populations in the results for either table 6 (spearman's rank-order correlation: .981, $\rho = .000$) or table 7 (spearman's rank-order correlation: .987, $\rho = .000$) Both of the probe populations here (Japanese and Egyin) are East Asians, one modern and one ancient, and thus geographically similar to Linzi. Therefore, the argument that the Linzi data is skewed purely by short sequence length appears to be incorrect, though sequence length is still an issue.

**Discussion**

First, we should give a brief discussion of some of the problems with this study not previously mentioned. The first issue is that this study only deals with maternal heritage. Analysis of the Y-chromosome could reveal differences in populations not revealed here, because some populations have experienced differential population histories varying by sex (e.g. due to long-distance migrations or matrilocal vs. patrilocal mating practices). One such possible example of this is that of Iceland, where according to Helgason et al. (2001) the original population consisted mainly of men from Scandinavia and women from the British Isles.

Further problems arise in this study from the large variation in sample sizes and sequence lengths (see table 1). These issues were dealt with as best as possible, with multiple runs and separate testing for the probe samples (the ancient samples) to try to maximize sequence lengths. The analysis of the effect of sequence length on this particular method does demonstrate that it is minimal. Otherwise, issues with available samples such as missing data limit the analyses' certainty in all cases, but hopefully using larger sample sizes and improving the quality of the source material can increase the accuracy of such genetic analysis. As far as sample size, for many populations all available data were used, and larger sample sizes will be available only with additional sample collection. Obviously, the results presented here should be taken generally, not as precise indicators of genetic affiliation.

Before we conclude, we should briefly discuss some of the problems of population-level genetic analysis. To look at very large sample sizes, examination at the population level may be the most efficient method, given a sufficient availability of computing power. However, the examination of genetic variation at the population level has its own set of problems, beginning with the simple problem of the definition of a "population." In addition, population similarity due to gene flow versus shared ancestry cannot be discerned with this method. Looking at possible paths of mutations, such as through haplogrouping, individual sequence-by-sequence comparison, or perhaps nested cladistic analysis within haplogroups (Templeton et al. 1995), might tease apart these issues. However, there are advantages in using direct nucleotide-to-nucleotide sequence comparisons between populations in large-scale studies, particularly with regard to higher resolution in the results (i.e. more detail to the distinctions), if technological and methodological issues can be overcome. Moreover, the above-mentioned methods can be employed in subsequent studies following a large-scale approach in order to further refine the results. Thus, we think that the methods and approach of this study are appropriate if interpreted with caution, particularly for the kind of large-scale population data examined in this case.

To conclude, genetic distances were estimated using a wide-angle lens, examining regional comparisons and extracting the best matches from each to create a total comparison ("free competition"). The goal was to eliminate the need to arbitrarily include or exclude populations from the overall comparison a priori, though obviously not every population in the world was included. However, it is felt by the authors that this method can produce more accurate comparisons than simply selecting populations based on preconceptions of population relationships or utilizing a single or small number of populations to represent whole regions. The problem with simply selecting arbitrary populations is highlighted by the original study on the Linzi material (Wang 2000), in which five random populations were chosen to represent Europe. This led to the incorrect attribution to the nearest relatives of the Linzi material being Icelander and Finnish (though they were possibly right about the Turkish comparison). No eastern Europeans or Iranians were included in that study, while these populations accompanied the Turkish as being closest to the Linzi material in this study, in which the Icelanders were actually not in the top ten. Obviously, there are still a number of "missing" populations in this analysis (such as the Tibetans, the Finns, the Russians, etc.). However, these methods are an improvement over arbitrary selection of populations for population comparison or the inclusion of only particular populations hypothesized to be related to the ancient population(s) based on linguistic or archaeological data. Of course, some regions of the world may not need such extensive analysis, but judging by the extremely variable populations and often bewildering history of Central Eurasia dating back at least to the late Paleolithic, it is quite applicable in this region.

The results suggest that there are definite differences in the genetic affinities between the ancient populations of Linzi in northern China and Egyin Gol in Mongolia. The Linzi material seems to bear a stronger affinity with Near Easterners and Europeans rather than with the present day populations of northern China, though there is a definite component of East and/or Southeast Asians within Linzi as well (as evidenced by haplogrouping, see below). We would suggest that rather than a "European-like population" in the ancient Linzi region, the Linzi material may be at least partially related to Indo-Iranians (a branch of Indo-European, though more precisely just "Iranian" by this time period), who were, during that period or at least shortly before it, probably inhabiting areas across Central Eurasia. More precisely, the Linzi population was quite possibly related to the Karsuk or Saka (putative Iranian groups who fit temporally and spatially), or also more distantly to the Andronovo, Afanasievo, Scythians, Sarmatians, or even the Sogdians. The Karsuk and Saka are the most likely given their existence in the 1$^{st}$ millennium BC in the central and possibly eastern parts of Central Eurasia, though these ethonyms are a little ambiguous and precise connections are not really possible. However, Harmatta (1992) has argued that early Iranian groups were spread across Central Eurasia from Eastern Europe to north China in the 1$^{st}$ millennium BC, and Askarov et al. (1992) have pointed out the existence of cist kurgan burials (with “Europoid” remains bearing some Mongoloid admixture, they suggest) in northwestern Mongolia in the same millennium. Although speculative, this line of reasoning fits in with other lines of evidence from archaeology and linguistics for the aforementioned changes in Chinese Bronze Age culture, the loan words in Old Chinese (Pulleyblank 1996, Kuzmina, 1998, Beckwith 2002, Di Cosmo 2002) and possibly sites like Zhukaigou and the Qijia culture (Linduff 1995), as well as evidence of Iranians on the steppe and possibly the Altai region at that time. The suggestion that actual European populations may have been in northern China at that time conflicts with general evidence of population movement on the steppe, which sees gradual movement of putative Indo-Iranians and Indo-Aryans throughout the steppe and associated areas in the 2nd and 1st millennia BC around Central Eurasia from the Indo-Aryans in India, the western Iranians on the Iranian plateau, and the Scythians and Sarmatians (and related groups) on the South Russian steppe and Eastern Europe. There is some evidence of them being on the Mongolic steppe (see Askarov et al. 1992), as well as evidence of their inhabitance of Xinjiang (such as Khotan) and possibly the Altai region (also the Tokharians, though they were not Indo-Iranian). Whether or not the Linzi site was populated by Iranian-like peoples (and whether these peoples came from the putative steppe Iranians to the west or from the possible Iranians of nearby ancient Xinjiang) is not clear from this study. However, this could explain the affinity between the Linzi site and the West. This would also fall in line with other evidence of admixture in populations in the region. Of course, the most difficult issue is that the early Iranians (and Indo-Iranians) were a linguistic group, and while perhaps bearing some biological affinities, the degree to which the supposed Iranian groups of Central Eurasia had a biological affinity is indeterminate.

As to why this study disagrees somewhat with previous results (Wang et al. 2000, Yao et al. 2003), there are several likely reasons. First, it should be noted that the results do agree with the two previous studies to some degree, in that the Linzi sample does appear to have some affinities to populations to the west as well as some populations of Southeast Asia, or at least southern China and Vietnam. The approach here cannot clearly account for significant admixture, as it rather weighs out the closest matches. Thus, the possibility exists that the Linzi population was a heavily admixed group containing elements from both the westerly populations as well as the southern Chinese and/or Southeast Asians. However, the argument by Yao et al. (2003) that the discrepancy is due purely to the shortness of the sequence length is not correct, as shown above (see tables 7 and 8). Yao et al. (2003) approached the issue by attempting to haplogroup the populations. However, due the shortness of the sequences, eight of the individuals were classified as unknown (about a quarter of the sample). Further, the six individuals classified as haplogroup B were in fact no different from CRS (Cambridge Reference sequence) in this segment (as a number of identified haplogroups from both Asia and Europe have no mutations in this particular segment). Also, six individuals classified as haplogroup B5A were found via a GenBank blast search to have near matches in both Asian and European populations, including Portuguese, Hungarians, Balkans, and Norse (all containing the two mutations, 16266A and 16274, though the Europeans also had an extra mutation here at 16258C, while all of the Asian matches except one lacked 16266A). The above discrepancies account for 20 out of the 34 individuals in Linzi. While the two proposed groups of individuals with haplogroups B and B5A would likely fall into B given their geographic location, we cannot simply assume that they are, or at least that they all are B (rather than H for instance). If we could simply assume individuals from Asia are all of "Asian" haplogroups (and likewise for other geographic locations), there would be no point in doing further research, since we already can assume the answer a priori.

This fact is further reinforced by a blast search analysis in GenBank of those individuals which were classified as unknown by Yao et al. 2003 and simply removed from the analysis (Linzi 7, 8, 10, 14, 21, 22, 24, 31). Linzi 21, 22, and 24 all contain a mutation at 16264 which was only evidenced in GenBank in an ancient Australian, though Linzi 21 also contained a mutation at 16355 which is found in a couple of modern Australian Aborigines as well as Scots, Georgians, Ossetians, Kazakhs and Norse (but never in tandem with 16264). Linzi 7, 8, and 14 contained mutations at 16231, 16256, 16270, and 16274. There were no exact matches with them, but the closest matches all came from the western part of Eurasia, including the populations of Adygeis (from the Caucasus), Syrians, Icelanders, Ossetians, Portuguese, Hungarians, Romanians, Serbians, Norse, Swedish and others. Linzi 10 and 31 were probably the most interesting, containing mutations at 16293 and 16311, with exact matches with Scottish, Greeks, Adygeis, Hungarians, Portuguese, Balkans, Slovakians, Estonians and others. All the exact matches were from western and central parts of Eurasia. The point here is to show that individuals were present at Linzi who likely were related to populations from western and/or central Eurasia. If this is the case, we can further suggest that the individuals who were automatically assumed by Yao et al. (2003) to be an Asian haplogroup, B for instance, may in fact potentially be something else, such as H (or at least some of them may be). The above evidence also highlights the problem of the presence of haplotypes in ancient populations which may be rare or nonexistent in modern populations, perhaps due to drift, coalescence, selective sweeps and other effects. While the above "unknown" haplotypes may not be part of the identified haplogroup paradigm, they certainly existed in the past, and in local populations may even have been prevalent to some degree. This problem is exacerbated when the analysis includes not only spatial variation, but temporal as well, as evolutionary forces can shift with time and situation.

The method here is designed to overcome the above problems. The haplotypes that could not be identified as a particular haplogroup would actually have a negligible effect on the results, as they would place Linzi equidistant from the various regional populations. It is the informative haplotypes and the mutations that comprise them, those which are rare or nonexistent in the other populations, which would make any given population have a relatively increased affinity with the probe population, in this case Linzi or Egyin. There are, of course, several individuals at Linzi who do appear to belong to haplogroups A, B, D, F, G, and M (all modern Asian haplogroups), thus explaining the results of Yao et al. (2003) as well as our own. However, as noted above, there are also a number of sequences that do appear to have an affinity to the west, which were thrown out by Yao et al. (2003) because they were not part of the known haplogroup paradigm, but likely explain the discrepancy between our results and theirs.

The Egyin Gol individuals appear to be definitely East Asians, at least maternally. The Egyin samples showed an affinity with northern East Asians, such as modern Mongolians, Japanese, northern Chinese populations (Shandong, Liaoning), and ethnic Han of Xinjiang. It is clear that the Egyin population was significantly different from the Linzi material from just a few centuries earlier. Of course, northern Mongolia to the Shandong region of China is actually some distance (well over a 1000 km). However, the historical connections of the Chinese to the Mongolic steppe in the first millennium BC, such as between the Hsiung-Nu and Han dynasty, show that the regions were in contact and had some degree of interaction (Watson 1961, Di Cosmo 2002). If the results from the Egyin Gol site can be duplicated by other finds from around the region, then there will be clear evidence that northern East Asians were the principle occupants of the area (including the steppe regions) by at least the rise of the Han dynasty in China (and perhaps the Qin dynasty or even earlier). Further correlation of the results of the Linzi site to other sites around the region dating to the middle of the first millennium BC and earlier (such as the genetic analysis by Ricaut et al. 2004 of an ancient individual from the Altai region) could provide evidence of a population shift in the region, depending of course on the degree to which populations like Linzi inhabited the region. It would also be useful to explore back into the Neolithic or even earlier to see whether the Linzi peoples were migrants to the region or descendents of earlier inhabitants. It should also be noted that if the attribution of the Egyin Gol site to the Hsiung-Nu is correct, then the results of the genetic analysis may suggest that the Hsiung-Nu were at least in part the ancestors of the later Mongolic and possibly Turkic peoples who would come to inhabit the steppe region in the first millennium AD. Further, if the evidence for a population shift can be corroborated and it is combined with the attribution of the Egyin Gol material to the Hsiung-Nu, then the Egyin Gol site may represent an element of some sort of genesis (though not necessarily the actual starting point), that would later result in the eruption of the Turkic and Mongol peoples from the Mongolic steppe and the Altai region of Central Eurasia. However, without correlation of the Linzi and Egyin Gol sites with other ancient sites from around the region, the evidence derived from these two sites will remain isolated cases.

**Acknowledgments**

We would like to thank Dr. Toomas Kivisild and Mait Metspalu of the Estonian Biocenter for providing their data for our use, and without whose help this study would not have been possible. We also thank Dr. Christopher Beckwith of the Central Eurasian Studies Department at Indiana University-Bloomington, Dr. Paul Jamison of the Anthropology Department at Indiana University-Bloomington, and Dr. Michael Wade of the Department of Biology at Indiana University-Bloomington for their review of a draft of this paper and thoughtful suggestions. We would also like to acknowledge Dick Repasky of the Research SP and UITS BioInformatics Support at Indiana University-Bloomington for his tireless efforts to help me resolve computing issues during this research, and Dr. F. Calafell for helping me locate sequence data.

**Literature Cited**

Al-Zahery, N., O. Semino, G. Benuzzi, C. Magri, G. Passarino, A. Torroni, and A.S. Santachiara-Benerecetti. 2003. Y-Chromosome and mtDNA Polymorphisms in Iraq, a Crossroad of the Early Human Dispersal and of Post-Neolithic Migrations. *Molec. Phyl. Evol.* 28(3): 458-472.

Anthony, D. 1998. The Opening of the Eurasian Steppe at 2000 BC. In: Mair, V. (ed) *The Bronze Age and Early Iron Age Peoples of Eastern Central Asia.* Washington, D.C.: The Institute for the Study of Man (with The University of Pennsylvania Museum Publications), 1: 94-113.

Beckwith, C. 2002. "The Sino-Tibetan Problem." In: Beckwith, C. (ed) *Medieval Tibeto-Burman Languages.* Leiden: EJ Brill, 113-157.

Benson D.A., I. Karsch-Mizrachi, D.J. Lipman, J. Ostell, B.A. Rapp, and D.L. Wheeler. 2000. GenBank. *Nucleic Acids Res.* 28(1): 15-18.

Bentley, J. 1993. *Old World Encounters*. New York: Oxford University Press.

Bertranpetit, J., J. Sala, F. Calafell, P.A. Underhill, P. Moral, and D. Comas. 1995. Human Mitochondrial DNA Variation and the Origin of the Basques. *Ann. Hum. Genet.* 59: 63-81.

Brakez, Z., E. Bosch, H. Izaabel, O. Akhayat, D. Comas, J. Bertranpetit, and F. Calafell. 2001. Human Mitochondrial DNA Sequence Variation in the Moroccan Population of the Souss Area. *Ann. Hum. Biol.* 28: 295-307.

Cavalli-Sforza, L.L., P. Menozzi, and A. Piazza. 1994. *The History and Geography of Human Genes*. Princeton, NJ: Princeton University Press.

Comas, D., F. Calafell, E. Mateu, Perez-Lezaun, A., and J. Bertranpetit. 1996. Geographic Variation in Human Mitochondrial DNA Control Region Sequence: the Population History of Turkey and its Relationship to the European Populations. *Mol. Biol. Evol.* 13(8): 1067-1077.

Comas, D., F. Calafell, E. Mateu, A. Perez-Lezaun, E. Bosch, R. Martinez-Arias, J. Clarimon, F. Facchini, G. Fiori, D. Luiselli, D. Pettener, and J. Bertranpetit. 1998. Trading Genes Along the Silk Road: mtDNA Sequences and the Origin of Central Asian Populations. *Am. J. Hum. Genet.* 63(6): 1824-1838.

Comas, D., F. Calafell, N. Bendukidze, L. Fananas, and J. Bertranpetit. 2000. Georgian and Kurd mtDNA Sequence Analysis Shows a Lack of Correlation Between Languages and Female Genetic Lineages. *Am. J. Phys. Anthropol.* 112(1): 5-16.

Di Cosmo, N. 2002. *Ancient China and Its Enemies: The Rise of Nomadic Power in East Asian History.* NewYork: Cambridge University Press.

Gresham, D., B. Morar., P.A. Underhill, G. Passarino, A.A. Lin, C. Wise, D. Angelicheva, F. Calafell, P.J. Oefner, P. Shen, I. Tournev, R. de Pablo, V. Kucinskas, A. Perez-Lezaun, E. Marushiakova, V. Popov, and L. Kalaydjieva. 2001. Origins and Divergence of the Roma (Gypsies). *Am. J. Hum. Genet.* 69(6): 1314-1331.

Golden, P.B. 1991. The Peoples of the South Russian Steppes. In: Sinor, D. (ed) *The Cambridge History of Early Inner Asia.* New York: Cambridge University Press.

Handt, O., S. Meyer, and A. von Haeseler. 1998. Compilation of Human mtDNA Control Region Sequences. *Nucleic Acids Res*. 26(1): 126-129.

Helgason, A., E. Hickey, S. Goodacre, B. Sykes, K. Steffanson, R. Ward,and V. Bosnes. 2001. mtDNA and the Islands of the North Atlantic: Estimating the Proportions of Norse and Gaelic Ancestry. *Am. J. Hum. Genet.* 68(3): 723-737.

Holsinger, K.E., and R.J. Mason-Gamer. 1996. Hierarchical Analysis of Nucleotide Diversity in Geographically Structured Populations. *Genetics* 142: 629-639.

Hudson, R.R., M. Slatkin, and W.P. Maddisson. 1992. Estimation of Levels of Gene Flow from DNA Sequence Data. *Genetics* 132: 583-589.

Imaizumi, K., T.J. Parsons, M. Yoshino, and M.M. Holland. 2002. A New Database of Mitochondrial DNA Hypervariable Regions I and II Sequences from Japanese Individuals. *Int. J. Leg. Med.* 116(2): 68-73.

Jorde, L.B., M.J. Bamshad, W.S. Watkins, R. Zenger, A.E. Fraley, P.A. Krakowiak, K.D. Carpenter, H. Soodyall, T. Jenkins, and A.R. Rogers. 1995. Origins and Affinities of Modern Humans – A Comparison of Mitochondrial and Nuclear Genetic Data. *Am. J. Hum. Genet.* 57(30): 523-538.

Keyser-Tracqui, C., E. Crubezy, and B. Ludes. 2003. Nuclear and Mitochondrial DNA Analysis of a 2,000-Year-Old Necropolis in the Egyin Gol Valley of Mongolia. *Am. J. Hum. Genet.* 73: 247-260.

Kivisild, T., M.J. Bamshad, K. Kaldma, M. Metspalu, E. Metspalu, E. Reidla, S. Laos, J. Parik, W.S. Watkins, M.E. Dixon, S.S. Papiha, S.S. Mastana, M.R. Mir, V. Ferak, and R. Villems. 1999. Deep common ancestry of Indian and western-Eurasian mitochondrial DNA lineages. *Curr. Biol.* 9(22): 1331-1334.

Kivisild, T., H.V. Tolk, J. Parik, Y. Wang, S.S. Papiha, H.J. Bandelt, R. Villems. 2002. The Emerging Limbs and Twigs of the East Asian mtDNA Tree. *Mol. Biol. Evol.* 19: 1737-1751.

Kivisild, T., S. Rootsi, M. Metspalu, S. Mastana, K. Kaldma, J. Parik, E. Metspalu, M. Adojaan, H.V. Tolk, V. Stepanov, M. Golge, E. Usanga, S.S. Papiha, C. Cinnioglu, R. King, L.L. Cavalli-

Sforza, P.A. Underhill, and R. Villems. 2003. The Genetic Heritage of the Earliest Settlers Persists Both in Indian Tribal and Caste Populations. *Am. J. Hum. Genet.* 72(2): 313-332.

Kong, Q.P., Y.G. Yao, M. Liu, S.P. Shen, C. Chen, C.L. Zhu, M.G. Palanichamy, and Y.P Zhang. 2003. Mitochondrial DNA Sequence Polymorphisms of Five Ethnic Populations from Northern China. *Hum Genet* 113(5): 391-405.

Kuzmina, E.E. 1998. Cultural Connections of the Tarim Basin People and Pastoralists of the Asian Steppes in the Bronze Age. In: Mair, V. (ed) *The Bronze Age and Early Iron Age Peoples of Eastern Central Asia.* Washington, D.C.: The Institute for the Study of Man (with The University of Pennsylvania Museum Publications, 1: 63-93.

Lattimore, O. 1951. *Inner Asian Frontiers of China.* New York: American Geographical Society.

Levine, M.A. 1999. Dereivka and the Problem of Horse Domestication. *Antiquity* 64: 727-740.

Linduff, K.M. 1995. Zhukaigou. *Antiquity* 69: 133-145.

Long, J.C., and R.A. Kittles. 2003. Human Genetic Diversity and the Nonexistence of Biological Races. *Hum. Biol.* 75(4): 449-471.

Lynch, M., and T.J. Crease 1990. The Analysis of Population Survey Data on DNA Sequence Variation. *Mol. Biol. Evol.* 7: 377-394.

Maddison, W.P., and D.R. Maddison 1989. Interactive Analysis of Phylogeny and Character Evolution Using the Computer Program MacClade. *Folia Primatol.* 53(1-4): 190-202.

Mair, V. 1995. Mummies of the Tarim Basin. *Archaeology* 48: 28-35.

Mair, V. (ed). 1998. *The Bronze Age and Early Iron Age Peoples of Eastern Central Asia.* Washington, D.C.: The Institute for the Study of Man (with The University of Pennsylvania Museum Publications).

Mallory, J.P. 1989. *In Search of Indo-Europeans: Language, Archaeology and Myth.* Slovenia: Thames and Hudson.

Nei, M. 1977. F-Statistics and Analysis of Gene Diversity in Subdivided Populations. *Ann. Hum. Genet.* 41: 225-233.

Nei, M. 1986. Definition and Estimation of Fixation Indexes. *Evolution* 40(3): 643-645.

Nei, M. 1987. *Molecular Evolutionary Genetics.* New York: Columbia University Press.

Narain, A.K. 1990. Indo-Europeans in Inner Asia. In: Sinor, D. (ed) *The Cambridge Early History of Early Inner Asia.* New York: Cambridge University Press.

Okladnikov, A.P. 1990. Inner Asia at the Dawn of History. In: Sinor, D. (ed) *The Cambridge Early History of Early Inner Asia.* New York: Cambridge University Press.

Oota, H., M. Naruya, and S. Ueda. 1999. Molecular Genetic Analysis of Remains of a 2000-Year-Old Human Population in China – And Its Relevance for the Origin of the Modern Japanese Population. *Am. J. Hum. Genet.* 64: 250-258.

Oota, H., W. Settheetham-Ishida, D. Tiwawech, T. Ishida, and M. Stoneking. 2001. Human mtDNA and Y-chromosome Variation is Correlated with Matrilocal Versus Patrilocal Residence. *Nat. Genet.* 29: 20-21.

Oota ,H., T. Kitano, F. Jin, I. Yuasa, L. Wang, S. Ueda, N. Saitou, and M. Stoneking. 2002. Extreme mtDNA Homogeneity in Continental Asian Populations. *Am. J. Phys. Anthropol.* 118(2): 146-153.

Parpola, A. 1998. Aryan Languages, Archaeological Cultures and Sinkiang, Where Did Proto-Iranian Come into Being and How Did It Spread. In: Mair, V. (ed) *The Bronze Age and Early Iron Age Peoples of Eastern Central Asia.* Washington, D.C.: The Institute for the Study of Man (with The University of Pennsylvania Museum Publications, 1: 114-147.

Plaza, S., F. Calafell, A. Helal, N. Bouzerna, G. Lefranc, J. Bertranpetit, and D. Comas. 2003. Joining the Pillars of Hercules: mtDNA Sequences Show a Multidirectional Gene Flow in the Western Mediterranean. *Ann. Hum. Genet.* 67: 312-328.

Praslov, N.D., V.N. Stanko, Z.A. Abromova, I.V. Sapozhnikov, and I.A. Borzijak. 1989. The Steppes in the Late Paleolithic. *Antiquity* 63: 784-792.

Pulleyblank, E.G. 1996. Early Contacts between Indo-Europeans and Chinese. *Int. Rev. Chinese Linguistics* 1(1): 1-25.

Redd, A.J. and M. Stoneking. 1999. Peopling of the Sahul: mtDNA Variation in Aboriginal Australian and Papua New Guinean Populations. *Am. J. Hum. Genet.* 65: 808-828.

Renfrew, C. 1987. *Archaeology and Language: The Puzzle of Indo-Europeans.* Cambridge, UK: Cambridge University Press.

Richards, M., H. CorteReal, P. Forster, V. Macauley, H. WilkinsonHerbots, A. Demaine, S. Papiha, R. Hedges, H.J. Bandelt, and B. Sykes. 1996. Paleolithic and Neolithic Lineages in the European Mitochondrial Gene Pool. *Am. J. Hum. Genet.* 59(1): 185-203.

Ricaut, F., C. Keyser-Tracqui, J. Bourgeois, E. Crubezy, amd B. Lydes. 2004. Genetic Analysis of a Scytho-Siberian Skeleton and Its Implications for Ancient Central Asian Migrations. *Hum Biol*. 76(1): 109-125.

Rozas, J., and R. Rozas. 1999. DnaSP Version 3: An Integrated Program for Molecular Population Genetics and Molecular Evolution Analysis. *Bioinformatics* 15: 174-175.

Sajantila, A., P. Lahermo, P. Anttinen, M. Lukka, P. Sistonen, M.L. Savontaus, P. Aula, L. Beckman, L. Tranebjaerg, T. Gedde-Dahl, L. Issel-Tarver, A. DiRienzo, and S. Paabo 1995. Genes and Languages in Europe: An Analysis of Mitochondrial Lineages. *Gen. Res.* 5: 42-52.

Sinor, D. (ed). 1990. *The Cambridge Early History of Early Inner Asia.* New York: Cambridge University Press.

Smouse, P.E., V.J. Vitzthum, and J.V. Neel. 1981. The Impact of Random and Lineal Fission on the Genetic Divergence of Small Human Groups: A Case Study Among the Yanomama. *Genetics* 98: 179-197.

Soong, D.L., H.S. Chang, S.L. Yoon, and B.L. Jung. 1997. Sequence Variation of Mitochondrial DNA Control Region in Koreans. *Forensic Sci Intl* 87: 99-116.

Templeton, A.R., E. Routman, and C.A. Phillips. 1995. Separating Population Structure from Population History: A Cladistic Analysis of the Geographical Distribution of Mitochondrial DNA Haplotypes in the Tiger Salamander, *Ambystoma tigrinum*. *Genetics* 140(2): 767-782.

Wang, L., H. Oota, N. Saitou, F. Jin, T. Matsushita, S. Ueda. 2000. Genetic Structure of a 2,500-Year-Old Human Population in China and Its Spatiotemporal Changes. *Mol. Biol. Evol.* 17(9): 1396-1400.

Watson, B. (trans). 1961. *Records of the Grand Historian of China (Ssuma Chien).* New York: Columbia University Press.

Weir, B.S., and C.C. Cockerham. 1984. Estimating F-Statistics for the Analysis of Population Structure. *Evolution* 38: 1358-1370.

Whitlock, M.C. 1994. Fission and Genetic Variance Among Populations: The Changing Demography of Forked Fungus Beetle Populations. *Am. Nat.* 143:820-829.

Yao, Y.G., L. Nie, H. Harpending, Y.X. Fu, Z.G. Yuan, and Y.P. Zhang. 2002a. Genetic Relationship of Chinese Ethnic Populations Revealed by mtDNA Sequence Diversity. *Am. J. Phys. Anthropol.* 118(1): 63-76.

Yao, Y.G., Q.P. Kong, H.J. Bandelt, T. Kivisild, Y.X. Fu, Z.G. Yuan, and Y.P. Zhang. 2002b. Phylogeographic Differentiation of Mitochondrial DNA in Han Chinese. *Am. J. Hum. Genet.* 70(3): 635-651.

Yao, Y.G., L. Nie, H. Harpending, Y.X. Fu, Z.G. Yuan, and Y.P. Zhang. 2002c. Genetic Relationships of Chinese Ethnic Populations Revealed by mtDNA Sequence Diversity. *Am. J. Hum. Genet.* 18(1): 63-80.

Yao, Y.G., Q.P. Kong, X.Y. Man, H.J. Bandelt, and Y.P. Zhang. 2003. Reconstructing the Evolutionary History of China: A Caveat About Inferences Drawn from Ancient DNA. *Mol. Biol. Evol.* 20(2):214-219.

Table 1: Population Data (in alphabetical order by region)

| Population | Haplotype Diversity | Nucleotide Diversity | Avg Num of Diff. | Variable Sites | Num Of Haplotypes | Num of Samples | Length Compared |
|---|---|---|---|---|---|---|---|
| Biaka[a] | 0.8900 | 0.02224 | 8.12 | 21 | 8 | 17 | 365 |
| Egyin[b] | 0.9740 | 0.01391 | 5.27 | 38 | 28 | 46 | 379 |
| Linzi[c] | 0.9360 | 0.01897 | 3.51 | 25 | 19 | 34 | 185 |
| **Europe** | | | | | | | |
| Armenian[d] | 0.9917 | 0.01586 | 5.71 | 109 | 130 | 192 | 360 |
| Basque[e] | * | * | * | * | * | * | * |
| Catalan[f] | 0.9180 | 0.01062 | 3.82 | 35 | 28 | 46 | 360 |
| Cuman[g] | 0.8910 | 0.00649 | 2.73 | 12 | 8 | 11 | 420 |
| Georgian[h] | 0.9900 | 0.01474 | 5.31 | 80 | 92 | 124 | 360 |
| German[i] | 0.9880 | 0.01723 | 4.72 | 60 | 74 | 108 | 274 |
| Hungarian[j] | 0.9870 | 0.01066 | 3.84 | 39 | 30 | 35 | 360 |
| Icelander[k] | 0.9734 | 0.01290 | 4.59 | 75 | 115 | 394 | 356 |
| Mari[l] | 0.9490 | 0.01150 | 4.13 | 18 | 10 | 13 | 359 |
| Moksha[l] | 0.9670 | 0.01255 | 4.42 | 26 | 15 | 21 | 352 |
| Moroccan[m] | 0.9610 | 0.01279 | 4.60 | 38 | 34 | 50 | 360 |
| RomB[n] | 0.9630 | 0.01297 | 4.67 | 30 | 15 | 20 | 360 |
| RomS[n] | 0.3250 | 0.00361 | 1.30 | 4 | 2 | 16 | 360 |
| Rom2S[n] | 0.7790 | 0.01440 | 5.18 | 32 | 16 | 57 | 360 |
| Saami[l] | 0.9000 | 0.01811 | 6.52 | 25 | 11 | 22 | 360 |
| Slovakians[d] | 0.9860 | 0.01274 | 4.58 | 71 | 89 | 129 | 352 |
| **S + SW Asia** | | | | | | | |
| Boqsa[o] | 0.9930 | 0.01482 | 5.31 | 33 | 17 | 18 | 358 |
| Iranian[p] | 0.9906 | 0.01514 | 5.69 | 135 | 281 | 435 | 376 |
| Iraqi[q] | 1.0000 | 0.01646 | 6.32 | 62 | 52 | 52 | 384 |
| Kashmir[r] | 0.9670 | 0.01627 | 5.86 | 32 | 14 | 18 | 360 |
| Kurds[s] | 0.9580 | 0.01194 | 4.29 | 40 | 22 | 29 | 359 |
| Lambadi[o] | 0.9830 | 0.01510 | 5.38 | 69 | 55 | 86 | 356 |
| Lobana[o] | 0.9790 | 0.01541 | 5.52 | 62 | 38 | 62 | 358 |
| Pakistan[o] | 1.0000 | 0.01543 | 5.56 | 18 | 9 | 9 | 360 |
| Parsi[h] | 0.9530 | 0.01275 | 4.58 | 47 | 29 | 55 | 359 |
| Pushtoon[p] | 0.9940 | 0.01647 | 5.91 | 53 | 32 | 36 | 359 |
| Tunisian[f] | 0.9900 | 0.01709 | 6.15 | 61 | 42 | 47 | 360 |
| Turks[t] | 0.9940 | 0.01494 | 5.38 | 56 | 40 | 45 | 360 |
| UttarPradesh[o] | 0.9920 | 0.01698 | 6.06 | 69 | 56 | 67 | 357 |
| **East Asia** | | | | | | | |
| Akha[u] | 0.9330 | 0.01684 | 5.09 | 33 | 24 | 91 | 302 |
| Ewenki[v] | 0.9560 | 0.01399 | 6.95 | 48 | 21 | 47 | 497 |
| Guangdong2[w] | 0.9950 | 0.01579 | 7.81 | 49 | 28 | 30 | 495 |
| Guangdong[x] | 0.9950 | 0.02165 | 7.71 | 68 | 62 | 70 | 356 |
| Indonesian[y] | 0.9660 | 0.02110 | 7.66 | 47 | 25 | 31 | 363 |
| Japanese[z] | 0.9847 | 0.01881 | 6.32 | 138 | 106 | 162 | 336 |
| KazakhXJ[p] | 0.9930 | 0.01360 | 6.76 | 46 | 27 | 30 | 497 |
| Kazakh[r] | 0.9900 | 0.01845 | 6.64 | 64 | 45 | 55 | 360 |
| KirghizHL[r] | 0.9840 | 0.01692 | 6.08 | 58 | 34 | 47 | 359 |
| KirghizLL[r] | 0.9950 | 0.01801 | 6.48 | 56 | 43 | 48 | 360 |
| Korean[z] | 0.9926 | 0.01519 | 5.74 | 136 | 207 | 306 | 378 |
| Liaoning[w] | 0.9980 | 0.01596 | 7.90 | 79 | 49 | 51 | 495 |
| Qidu[c] | 0.9900 | 0.02524 | 4.67 | 43 | 41 | 50 | 185 |
| Mongolian[v] | 0.9890 | 0.01415 | 7.02 | 64 | 38 | 48 | 496 |
| Shandong[w] | 0.9950 | 0.01445 | 7.18 | 67 | 44 | 49 | 497 |
| UighurXJ[p] | 0.9950 | 0.01291 | 6.42 | 64 | 41 | 45 | 497 |
| Uighur[r] | 0.9930 | 0.01641 | 5.91 | 63 | 46 | 55 | 360 |
| Vietnamese[aa] | 0.9930 | 0.02538 | 6.93 | 55 | 31 | 35 | 273 |
| Wuhan[w] | 1.0000 | 0.01618 | 8.03 | 63 | 42 | 42 | 497 |
| XinjiangHan[w] | 0.9950 | 0.01440 | 7.14 | 62 | 43 | 47 | 496 |
| Yunnan[w] | 0.9930 | 0.01699 | 8.43 | 56 | 40 | 43 | 497 |
| Total (w/ Linzi/Qidu) | 0.9727 | 0.02344 | 3.40 | 132 | 855 | 3703 | 145 |
| Total ((w/o Linzi/Qidu | 0.9849 | 0.01888 | 4.10 | 174 | 1084 | 3619 | 217 |
| Averages | 0.9594 | 0.01545 | 5.73 | 55 | 48 | 71 | 377 |

[a]Jorde et al. 1995

[b]Keyser-Tracqui et al. 2003

[c]Wang et al. 2000

[d]Unpublished GenBank, Metspalu et al.

[e]Bertranpetit et al. 1995

[f]Plaza et al. 2003

[g]Unpublished GenBank, Szabo et al.

[h]Unpublished GenBank, Riedla et al.

[i]Richards et al. 1996

[j]Unpublished GenBank, Kalmar et el

[k]Helgason et al. 2001

[l]Sajantila et al. 1995

[m]Brakez et al. 2001

[n]Unpublished GenBank, Kaldma et al.

[o]Unpublished GenBank, Kivisild et al.

[p]Personal communication from T. Kivisild and M. Metspalu

[q]Al-Zahery et al. 2003

[r]Comas et al. 1998

[s]Comas et al. 2000

[t]Comas et al. 1996

[u]Oota et al. 2001

[v]Kong et al. 2003

[w]Yao 2002b

[x]Kivisild et al. 2002

[y]Redd and Stoneking 1999

[z]Imaizumi et al. 2002

[y]Lee et al. 1997

[z]Oota et al. 2002

Table 2: European Regional Fst Comparison

| | Linzi | | Egyin |
|---|---|---|---|
| Hungarian | 0.03204 | RomB | 0.14389 |
| Armenian | 0.04060 | Moroccan | 0.14964 |
| Catalan | 0.04108 | Armenian | 0.15801 |
| Slovakians | 0.04427 | Catalan | 0.16256 |
| Basque | 0.04532 | Georgian | 0.16335 |
| Icelander | 0.04914 | Cuman | 0.16674 |
| Moroccan | 0.04944 | Hungarian | 0.18061 |
| Mari | 0.05265 | Slovakians | 0.18550 |
| RomB | 0.05333 | Icelander | 0.18573 |
| Georgian | 0.05517 | Basque | 0.18801 |
| Cuman | 0.05701 | German | 0.19103 |
| German | 0.05873 | Rom2S | 0.19245 |
| Moksha | 0.06409 | Saami | 0.19978 |
| Saami | 0.10972 | Moksha | 0.20307 |
| Rom2S | 0.15153 | Mari | 0.23218 |
| Egyin | 0.16790 | RomS | 0.41694 |
| Biaka | 0.40316 | Biaka | 0.48456 |
| RomS | 0.40981 | | |

Table 3: South and Southwest Asia Regional Fst Comparison

| | Linzi | | Egyin |
|---|---|---|---|
| Turks | 0.03210 | Lambadi | 0.0502 |
| Iranian | 0.03736 | Lobana | 0.07141 |
| Kashmir | 0.04228 | Boqsa | 0.08162 |
| Iraqi | 0.04303 | Tunisian | 0.08182 |
| Pushtoon | 0.04870 | Parsi | 0.08868 |
| Tunisian | 0.05097 | Pushtoon | 0.08868 |
| Kurds | 0.05411 | UttarPradesh | 0.09845 |
| Pakistan | 0.05504 | Pakistan | 0.10115 |
| UttarPradesh | 0.05859 | Kashmir | 0.10537 |
| Lambadi | 0.08922 | Turks | 0.12799 |
| Parsi | 0.09469 | Iranian | 0.15425 |
| Lobana | 0.09513 | Iraqi | 0.16438 |
| Boqsa | 0.12558 | Kurds | 0.17653 |
| Egyin | 0.16310 | Biaka | 0.42833 |
| Biaka | 0.40777 | | |

Table 4: East and Central Asia Regional Fst comparison

| | Linzi | | Egyin |
|---|---|---|---|
| Vietnamese | 0.03788 | XinjiangHan | 0.01473 |
| Uighur | 0.04740 | Japanese | 0.01568 |
| Guangdong2 | 0.05699 | Liaoning | 0.01688 |
| Yunnan | 0.05935 | Shandong | 0.01853 |
| KirghizHL | 0.06037 | Ewenki | 0.02006 |
| Kazakh | 0.06932 | Mongolian | 0.02369 |
| Wuhan | 0.07937 | KirhgizLL | 0.02415 |
| UighurXJ | 0.08150 | Korean | 0.02612 |
| Guangdong | 0.08893 | UighurXJ | 0.03033 |
| XinjiangHan | 0.09207 | Kazakh | 0.03681 |
| Indonesian | 0.09692 | Guangdong2 | 0.04247 |
| KazakhXJ | 0.09806 | Wuhan | 0.04659 |
| Liaoning | 0.10004 | KirghizHL | 0.04690 |
| Ewenki | 0.10533 | Akha | 0.05675 |
| KirghizLL | 0.12497 | Uighur | 0.05808 |
| Qidu | 0.12570 | KazakhXJ | 0.05823 |
| Shandong | 0.12636 | Yunnan | 0.07254 |
| Korean | 0.12977 | Guangdong | 0.09830 |
| Mongolian | 0.13279 | Indonesian | 0.09948 |
| Akha | 0.13417 | Vietnamese | 0.11113 |
| Japanese | 0.14389 | Biaka | 0.43003 |
| Egyin | 0.16471 | | |
| Biaka | 0.40158 | | |

Table 5: Total Fst Comparison

| | Linzi | | Egyin |
|---|---|---|---|
| Hungarian | 0.03084 | XinjiangHan | 0.01289 |
| Turks | 0.03201 | Mongolian | 0.01586 |
| Iranian | 0.03765 | Shandong | 0.01630 |
| Vietnamese | 0.03776 | Japanese | 0.01755 |
| Armenian | 0.04077 | Liaoning | 0.01857 |
| Catalan | 0.04095 | KirghizLL | 0.01972 |
| Slovakians | 0.04302 | Korean | 0.02271 |
| Iraqi | 0.04334 | Ewenki | 0.02288 |
| Basque | 0.04413 | UighurXJ | 0.03123 |
| Kashmir | 0.04436 | Kazakh | 0.03355 |
| Uighur | 0.04740 | Lambadi | 0.05030 |
| Icelander | 0.04777 | Lobana | 0.07208 |
| Moroccan | 0.04898 | Boqsa | 0.08292 |
| Pushtoon | 0.04900 | Tunisian | 0.08344 |
| Pakistan | 0.05035 | Parsi | 0.08908 |
| Tunisian | 0.05161 | Pushtoon | 0.08955 |
| Mari | 0.05206 | UttarPradesh | 0.09998 |
| Kurds | 0.05411 | Pakistan | 0.10157 |
| RomB | 0.05575 | RomB | 0.13780 |
| Guangdong | 0.05668 | Moroccan | 0.15186 |
| Yunnan | 0.05935 | Cuman | 0.15466 |
| KirghizHL | 0.06048 | Armenian | 0.16654 |
| Kazakh | 0.06932 | Georgian | 0.16772 |
| Wuhan | 0.07937 | Hungarian | 0.17357 |
| UighurXJ | 0.08106 | Catalan | 0.17485 |
| Guangdong2 | 0.08803 | Slovakians | 0.18742 |
| XinjiangHan | 0.09163 | Icelander | 0.19261 |
| Egyin | 0.16390 | Biaka | 0.44516 |
| Biaka | 0.40158 | | |

Table 6: Comparison of Egyin and Egyin Limited (EgLim)

| | Egyin | | EgLim |
|---|---|---|---|
| XinjiangHan | 0.01289 | Japanese | 0.01330 |
| Mongolian | 0.01586 | Qidu | 0.01373 |
| Shandong | 0.01630 | XinjiangHan | 0.01522 |
| Japanese | 0.01755 | Shandong | 0.01637 |
| Liaoning | 0.01857 | Liaoning | 0.01839 |
| KirghizLL | 0.01972 | Mongolian | 0.01958 |
| Korean | 0.02271 | Ewenki | 0.02240 |
| Ewenki | 0.02288 | KirghizLL | 0.02490 |
| UighurXJ | 0.03123 | Korean | 0.02503 |
| Kazakh | 0.03355 | UighurXJ | 0.03475 |
| Lambadi | 0.05030 | Kazakh | 0.04085 |
| Lobana | 0.07208 | Lambadi | 0.05566 |
| Boqsa | 0.08292 | Parsi | 0.08993 |
| Tunisian | 0.08344 | Lobana | 0.09028 |
| Parsi | 0.08908 | Tunisian | 0.10539 |
| Pushtoon | 0.08955 | Pushtoon | 0.10622 |
| UttarPrad | 0.09998 | Pakistan | 0.10656 |
| Pakistan | 0.10157 | UttarPrad | 0.10725 |
| RomB | 0.13780 | Boqsa | 0.10910 |
| Moroccan | 0.15186 | RomB | 0.16122 |
| Cuman | 0.15466 | Linzi | 0.16390 |
| Armenian | 0.16654 | Moroccan | 0.17746 |
| Georgian | 0.16772 | Cuman | 0.18438 |
| Hungarian | 0.17357 | Armenian | 0.19834 |
| Catalan | 0.17485 | Georgian | 0.20749 |
| Slovakians | 0.18742 | Catalan | 0.20935 |
| Icelander | 0.19261 | Hungarian | 0.21305 |
| Biaka | 0.44516 | Slovakians | 0.22333 |
| | | Icelander | 0.22652 |
| | | Biaka | 0.42503 |

Table 7: Comparison of Japanese and Japanese Limited (JpLim)

| | Japanese | | JapLim |
|---|---|---|---|
| Liaoning | 0.00307 | Liaoning | 0.00373 |
| XinjiangHan | 0.01073 | XinjiangHan | 0.01182 |
| Guangdong2 | 0.01416 | Egyin | 0.01264 |
| Egyin | 0.02443 | Uighur | 0.02007 |
| Wuhan | 0.02606 | Guangdong2 | 0.02446 |
| Kazakh | 0.03093 | Kazakh | 0.02942 |
| Yunnan | 0.03239 | Wuhan | 0.03839 |
| Uighur | 0.03735 | Yunnan | 0.04133 |
| KirghizHL | 0.03861 | KirghizHL | 0.04228 |
| UighurXJ | 0.04405 | UighurXJ | 0.04591 |
| Guangdong | 0.05123 | Guangdong | 0.07404 |
| Tunisian | 0.06366 | Tunisian | 0.07680 |
| Kashmir | 0.07581 | Pushtoon | 0.08314 |
| Pushtoon | 0.08703 | Pakistan | 0.08572 |
| Pakistan | 0.10333 | Vietnamese | 0.09493 |
| RomB | 0.11529 | Kashmir | 0.10624 |
| Turks | 0.11568 | Turks | 0.11663 |
| Moroccan | 0.13088 | RomB | 0.13226 |
| Iranian | 0.13130 | Moroccan | 0.14273 |
| Armenian | 0.13874 | Linzi | 0.14407 |
| Catalan | 0.14550 | Iranian | 0.14873 |
| Kurds | 0.14816 | Iraqi | 0.15740 |
| Iraqi | 0.15044 | Armenian | 0.16736 |
| Hungarian | 0.15873 | Catalan | 0.17449 |
| Basque | 0.15943 | Kurds | 0.17596 |
| Slovakians | 0.16734 | Hungarian | 0.17949 |
| Icelander | 0.16840 | Slovakians | 0.18962 |
| Mari | 0.21979 | Icelander | 0.19481 |
| Biaka | 0.41035 | Basque | 0.19790 |
| | | Mari | 0.24977 |
| | | Biaka | 0.38679 |